# From manuscript catalogues to a handbook of Syriac literature: Modeling an infrastructure for Syriaca.org

**Nathan P. Gibson[1*], David A. Michelson[2], Daniel L. Schwartz[3]**

1 Vanderbilt University, USA

2 Vanderbilt University, USA

3 Texas A&M University, USA

*Corresponding author: Nathan Gibson editors@syriaca.org

**Abstract**
Despite increasing interest in Syriac studies and growing digital availability of Syriac texts, there is currently no up-to-date infrastructure for discovering, identifying, classifying, and referencing works of Syriac literature. The standard reference work (Baumstark's *Geschichte*) is over ninety years old, and the perhaps 20,000 Syriac manuscripts extant worldwide can be accessed only through disparate catalogues and databases. The present article proposes a tentative data model for Syriaca.org's *New Handbook of Syriac Literature*, an open-access digital publication that will serve as both an authority file for Syriac works and a guide to accessing their manuscript representations, editions, and translations. The authors hope that by publishing a draft data model they can receive feedback and incorporate suggestions into the next stage of the project.

## INTRODUCTION

### What is Syriac?
The Syriac language is a dialect of Aramaic that had its origins around the first century C.E. near where the modern borders of Syria, Iraq, and Turkey converge. By the time of the Arab-Muslim conquests in the seventh century, this language was used as a *lingua franca* in travel, trade, and religious culture from the eastern Mediterranean to Central Asia. Even after Arabic became the common language of the Fertile Crescent in the ensuing centuries, Syriac literature and networks of Syriac-speaking scholars served as a cultural bridge between the Greek world of Byzantium and the Arab courts of the Islamic states. Minority communities (especially Christian ones) have been producing Syriac literature up until the present day. Syriac texts today comprise one of the richest troves of literary and historical material from the late antique and medieval Middle East [see Briquel Chatonnet and Debié, 2015, 1; Brock, 2006, 1-2].

## Why a New Handbook of Syriac Literature?

Currently, there is no resource that attempts to provide reference information and standardized titles or identifiers for all Syriac works. Researchers generally have two options for finding Syriac texts: (1) consulting finding aids such as catalogues or journal articles for each manuscript that might contain the text ([Desreumaux and Briquel-Chatonnet, 1991] list over 800 finding aids, many not available as online databases) or (2) using one of several histories of Syriac literature. The most comprehensive and widely used one is almost a century old: [Baumstark, 1922; see Tannous et al, 2012-2016]. Baumstark's work, which was based on his impressive knowledge of Syriac manuscripts, was not comprehensive at the time, let alone now, when old manuscripts have moved, new manuscripts have surfaced, and debates in the secondary literature have reshaped our understanding of authors and the works they wrote [compare McCollum, 2015, 66]. Moreover, Baumstark's complex prose and dense abbreviations make his book a tool for advanced users only. With support of the National Endowment for the Humanities and the International Balzan Prize Foundation, Syriaca.org is preparing a new, highly accessible guide to Syriac literature that enables users to identify, locate, and cite any Syriac work.

## Why a digital tool?

Besides the typical advantages that digital publication offers, such as increased accessibility, there are three specific ways that a handbook of Syriac literature can benefit from digital humanities methods: (1) interdisciplinary connections, (2) cultural preservation, and (3) infrastructure creation.

First, the types of works represented in Syriac texts are vast, ranging from tomb inscriptions of only a few words to voluminous world chronicles; from folk tales to scholarly treatises; and from routine letters and technical jurisprudence to lofty liturgical poetry [see Brock, 1997]. Many of these works were originally composed in Syriac, but a good number were translated or adapted from texts in Greek or other languages. In fact, Syriac translators played a key role in the golden age of Arabic philosophy, which in turn was one of the crucial factors leading to the renewed interest in Greek philosophical works during the Italian Renaissance [Brock, 1997, 11]. As is apparent, Syriac literature has relevance for a wide variety of scholars, among them historians of many kinds, archaeologists, philosophers, and theologians. The breadth and depth of this set of texts crosses regional, chronological, and disciplinary boundaries. Linked open data approaches allow a digital handbook to interface with a multitude of fields in ways that were previously inconceivable.

Second, political and social events have led to the near extinction of the Syriac language and its siblings (various dialects of Neo-Aramaic) [see Briquel Chatonnet and Debié, 2015, 3]. The conflicts of recent years in the Middle East have struck a hard blow to the Syriac communities, which were already in crisis. Apart from a massive effort at cultural preservation, it is difficult to predict whether in a few decades the readers (not to mention writers) of Syriac literature will be anyone other than scholars of archaic languages. A digital resource can encourage heritage communities not just to access cultural archives, but also to engage with them and contribute to them.

Third, creating a "born-digital" standard reference work will allow Syriac studies to "leapfrog" certain developmental steps found in other subsets of ancient history. Whereas some ancient languages, such as Greek, Latin, and Hebrew, have seen new waves of reference works published in print over the last few decades that are now being migrated into a digital infrastructure, many standard Syriac reference works are 100-200 years old. As debilitating as this has been for Syriac scholarship, it also suggests an opportunity: new Syriac reference works that are published digitally using robust frameworks and open licenses can rapidly reorient the field toward a digital future——one with fewer copyright limitations and less "digitization baggage" (such as the overhead costs of reconciling print and digital resources) than those of comparatively more developed fields. The potential here is to create an infrastructure that will serve the field for years to come.

**What are our goals?**
The following goals for *The New Handbook of Syriac Literature* (*NHSL*) will determine our strategies for creating it:

- **Comprehensiveness:** Due to constraints of time and funding, the first edition will not provide exhaustive coverage of all Syriac works, nevertheless we want to adopt strategies that will scale efficiently over time so that the work can be continued until it approaches complete coverage.
- **Accessibility:** *NHSL* will be released under a Creative Commons Attribution 3.0 Unported license [http1, 2016]. All data, except user suggestions and preferences, will be accessible without login.
- **Usability:** The human interface for *NHSL* should be attractive and easy to navigate without technical expertise.
- **Extensibility:** The format of the published dataset should allow for later additions and revisions by the Syriaca.org editorial team and for suggestions from the community of users.
- **Interoperability:** The published dataset should be machine-processable using widely accepted technologies and should provide the necessary entry points and documentation for linking with other projects. It should also be able to incorporate data in open formats coming from other projects.
- **Breadth of Audience:** *NHSL* should serve research, archival, learning, and heritage communities. This will require providing good metadata with clear human-readable explanations, interfacing with a variety of types of websites, and making data available in diverse formats.

## I CONCEPTUAL MODEL AND ENTITIES

### 1.1 Textual entities and URIs
At the core of developing *The New Handbook of Syriac Literature* is our plan to create records with stable URIs for every work, manuscript, and publication relevant to Syriac literature for which we have sufficient and reliable source material. By **work** we mean an abstract or conceptual entity consisting of all written texts that have the same intellectual source.[1] (See below for a comparison of this concept of work with that of FRBR.) We use the term **manuscript** to refer to any kind of object bearing handwritten text. A **publication**, or bibliography item, is any discretely titled textual item or compilation of items which our sources reference or which we deem useful to the reader. The URI formats for these three entities are http://syriaca.org/work/{\d+}, http://syriaca.org/manuscript/{\d+}, and http://syriaca.org/bibl/{\d+}, respectively.

Properly mapping these relationships among works, manuscripts, and publications is essential to providing a useful reference volume for Syriac literature: users need to be able to view a single work record that displays the manuscripts and editions or translations where that work can be consulted. At present, it can take hours or days of searching catalogues and tracking down citations to identify the manuscripts and editions of a Syriac work. Thus, the pragmatic value of having these listed and linked in one place is very high. For *NHSL*, the work record will function as a kind of hub bringing together this information.

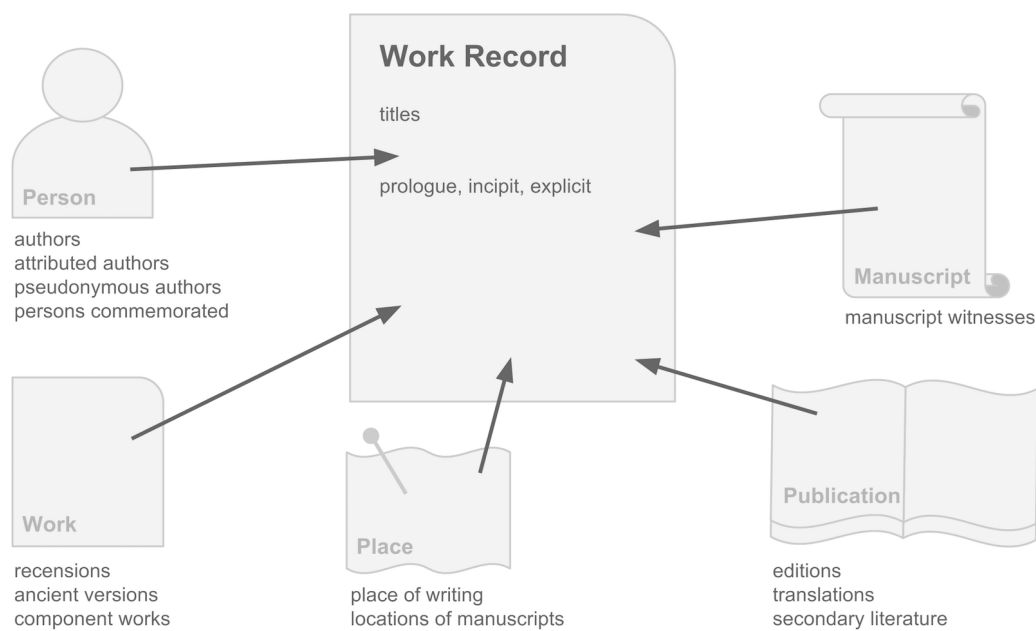

Figure 1. A conceptual diagram of the work record as a hub for literary information, with sample (non-comprehensive) relationships to other entities.

Yet these relationships also raise some conceptual and practical difficulties. For example, how does one decide whether a text whose manuscript witnesses differ considerably should be considered one work with one URI or multiple works with multiple URIs? Should an ancient translation (or "version") of a text be considered the same work or a different one, especially if it significantly abridges or paraphrases it?

---

[1] This is essentially identical to the LAWD definition of a "Conceptual Work" [http2, 2016] and quite similar to the BIBFRAME definition of a "Work" [http3, 2016].

## 1.2 Textual relationships and FRBR

Most of the discussion about distinguishing the intellectual contents of works from the physical items that represent or embody those works centers on the model produced by the study called Functional Requirements for Bibliographic Records (FRBR) [http4, 2016].

FRBR defines four types in its "Group 1" entities:
- work: "a distinct intellectual or artistic creation,"
- expression: "the intellectual or artistic realization of a *work* in the form of alpha-numeric, musical, or choreographic notation, sound, image, object, movement, etc., or any combination of such forms,"
- manifestation: "the physical embodiment of an *expression* of a *work*," and
- item: "a single exemplar of a *manifestation*" [http5, 2016].

What would it look like to implement a FRBR model for the entities we want to describe in *NHSL* (see figure 2)? Syriaca.org "publications" clearly correspond to FRBR manifestations. Each manuscript, however, would need to be described using multiple FRBR entities: an item, a manifestation, and perhaps even an expression, to the extent that its text varies from other manuscripts. Syriaca.org "works" could often correspond to FRBR works, but recensions and translations (ancient or modern) could arguably constitute individual expressions.

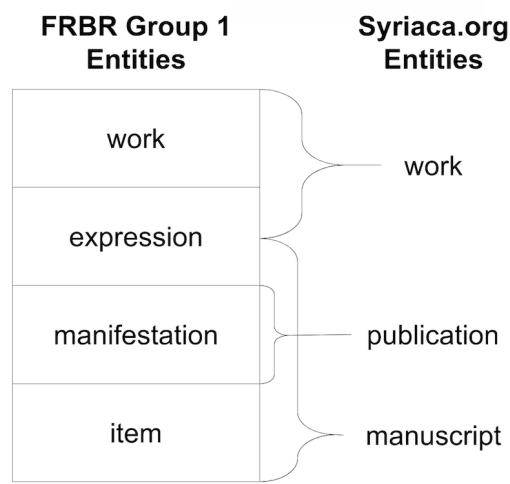

Figure 2. FRBR Group 1 Entities compared with Syriaca.org entities.

The end result would be (1) creating a number of entities and relationships that are unnecessary for our purposes and (2) distinguishing between types of entities in ways that do not serve our users. As an example of (1), consider the case of a manuscript witness to a text that has minor textual variations from other manuscripts. Would it serve any useful purpose to mint three separate URIs (expression, manifestation, and item) for a single manuscript? Should one insist that references to the manuscript's textual variants cite its expression URI, while accounts of its foliation mention its manifestation URI, and descriptions of it as a physical object refer to its item URI? Regarding (2), take for example the case of a Greek work that has multiple Syriac versions that resemble it more or less closely. Should we create expression entities for the original Greek text as well as for each Syriac version? According to the FRBR study, translations are different expressions of a single work, but modifications of a work are considered new works if they involve "a significant degree of independent intellectual or artistic effort" [http5, 2016]. If one Syriac version is a paraphrastic translation of the Greek text whereas another has numerous interpolations, should the first be considered an expression and the second a separate work? Such a distinction would require us to make editorial decisions that do not benefit our users. Why cannot both Syriac versions simply be works that have the Greek work as their source? We want to be able to show our users a page for each work that shows all of the texts that have that work as their source, with their editions and manuscripts. Distinguishing between works and expressions simply makes this more difficult.

In fact, post-FRBR linked-data models developed in library science have adopted the practice of grouping bibliographic records under Works without the Expression entity (see figure 3 below). BIBFRAME [http6, 2016], for example, the linked-data vocabulary developed by the Library of Congress, defines a number of Work-Work relationships (such as bf:translation/bf:translationOf) that should be Work-Expression relationships according to the FRBR model. Thus, we believe *NHSL* may be able to use a simplified Work model without diminishing its value to users or its compatibility with potential library partners.

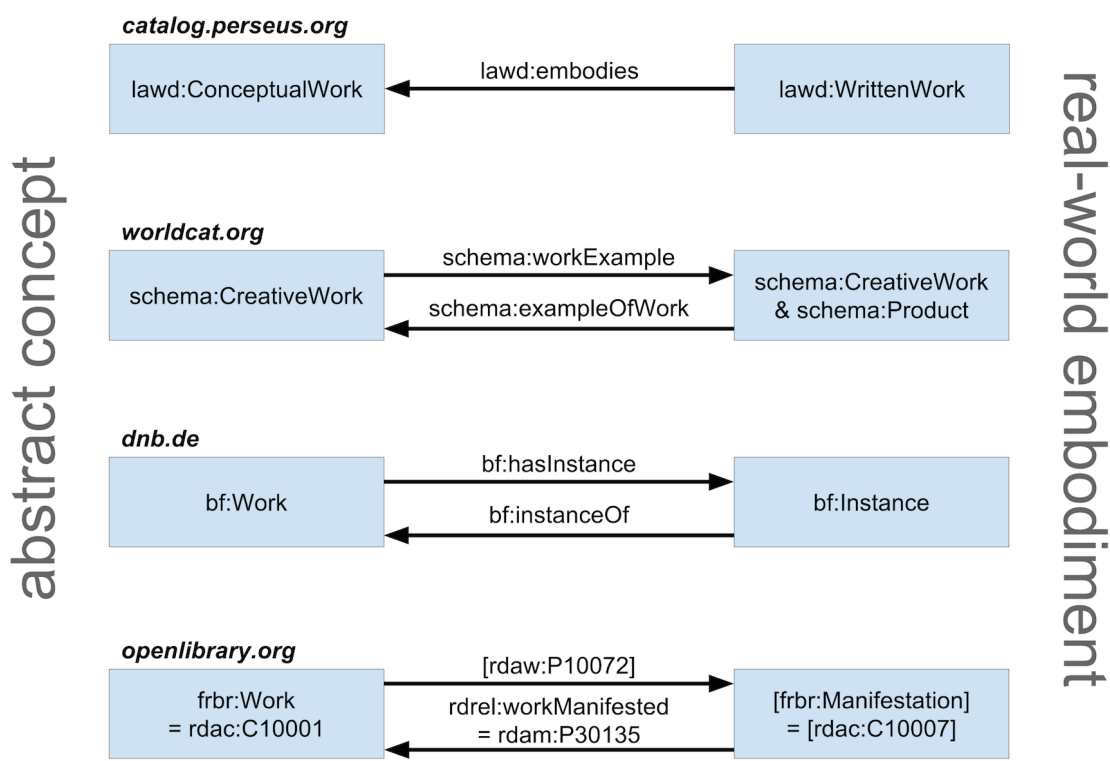

Figure 3. Examples of linked-data models relating abstract works to their real-world embodiments. Note: This comparison is intended to show only functional similarities (not formal equivalences) among the above classes and properties.

The solutions we are exploring for translations and recensions are pragmatic ones. When we have data for multiple manuscripts or publications relating to a specific version or recension of a work, we may group these under a separate work entity with its own URI and include a relationship to the source work entity denoting the specific type of connection (for example, syriaca:hasVersion or syriaca:hasRecension). When we have data for only one manuscript or publication relating to a particular version or recension (and therefore do not need to show our users a separate results list), we may link the main work entity directly to this manuscript or publication using a relationship property that corresponds to the type of relationship that separate works would have to each other (for example, syriaca:hasEmbodiedVersion or syriaca:hasEmbodiedRecension). Such a relationship would imply the existence of an intermediate work and would allow us or our partners to expand the relationship later, if needed, without creating unnecessary work entities at the outset. These options are still under consideration. We are actively looking for models from other projects that we can replicate, and we would welcome feedback.

### 1.3 Relationships to other entities

In addition, the textual entities will link to at least two other types of entities contained in Syriaca.org's datasets: persons and places. Relationships to persons will necessarily include the works' authors, contained in *A Guide to Syriac Authors* [Michelson et al, forthcoming 2016], and saints commemorated in hagiographical works, described in *Qadishē: Guide to the Syriac Saints* [Saint-Laurent and Michelson, forthcoming 2016]. The links from works to places (such as place of writing, find location, repository, or publication place) will be to records in *The Syriac Gazetteer* [Carlson and Michelson, 2014]. In addition, an ongoing prosopographical project based on Syriac texts (*SPEAR: Syriac Persons, Events, and Relations* [Schwartz, forthcoming]) will also connect Syriac works to persons and places.

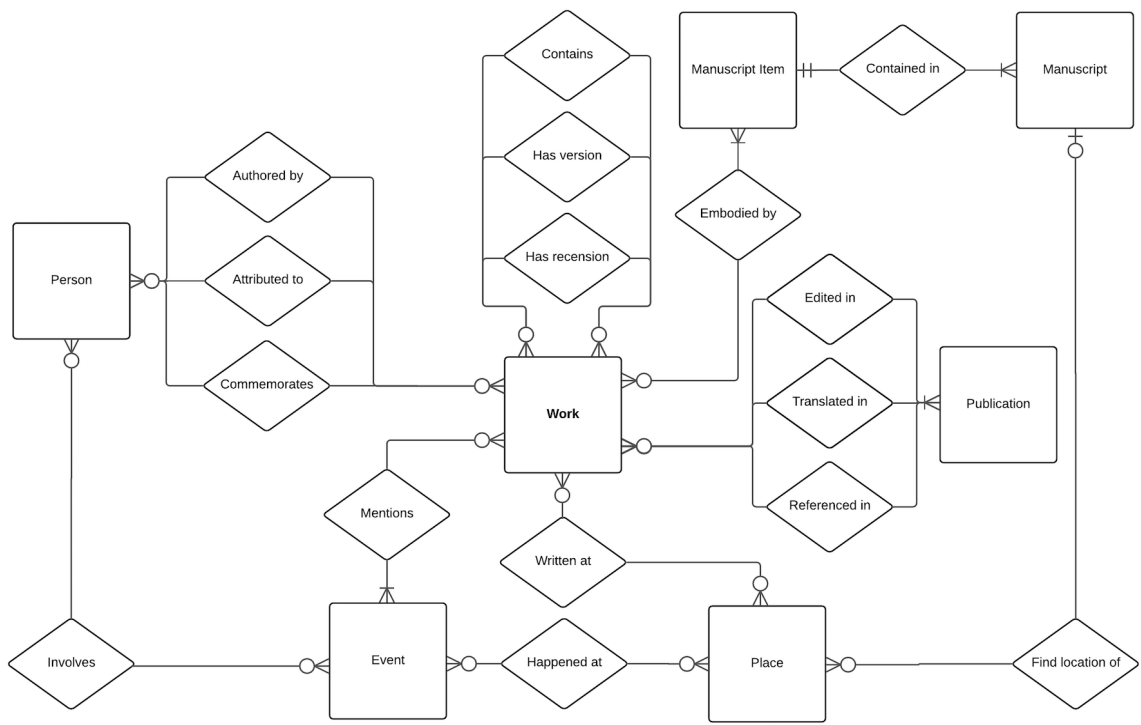

Figure 4. Entity Relationship Diagram for *NHSL* from the perspective of the work entity.

## 1.4 Preliminary classification of Syriac works

So far, Syriac literature lacks any kind of thorough classification system. We hope that *NHSL* will make this more achievable. While we would like to classify Syriac works according to formal and stylistic genres, this will need to wait until we have combed through large amounts of data for *NHSL*. For now, we propose labeling the works we are cataloguing using the following preliminary subject taxonomy adapted from three widely used works on Syriac literature [Wright, 1870-1872; Brock, 1997; Duval, 1907]. This taxonomy will be reviewed and revised once a draft of *NHSL* is ready.

1. Bible
   a. Individual Biblical Books
   b. Collected Biblical Books
   c. Lectionaries
   d. Psalters
2. Works of Theology and Biblical Interpretation/Exposition
   a. Bible Commentaries
   b. Poems on Biblical Events
   c. Bible Vocalization (*Masora*)
   d. Scholia
   e. Questions and Answers
   f. Other Theological Works
3. Liturgical Works
   a. Missals
   b. Daily Offices

      c.  Ecclesiastical Office Books
      d.  Collections of Choral Services, Homilies, and Hymns
      e.  Prayers
      f.  Services for Baptisms, Funerals, and other Life Events
4.  Works on Spiritual and Ascetic Disciplines
5.  Hagiographical Works
      a.  Martyr Acts
      b.  Lives of Saints and Ascetics
      c.  Poems on Saints and Ascetics
      d.  Other Hagiographical Works
6.  Apologetic and Heresiological Works
7.  Legal and Ecclesiastical Works
      a.  Accounts and Records of Ecclesiastical Councils
      b.  Ecclesiastical Canons/Works of Canon Law
      c.  Works of Civil Law
8.  Histories
      a.  Universal Histories
      b.  Particular Histories
9.  Philosophical Works
      a.  Works on Logic
      b.  Works on Ethics
10. Linguistic Works
      a.  Grammatical Works
      b.  Lexicographic Works
      c.  Works on Rhetoric and Poetics
11. Works on the Natural and Therapeutic Sciences
      a.  Works on Medicine
      b.  Works on Agriculture
      c.  Works on Chemistry
      d.  Works on Astronomy, Cosmography, and Geography
12. Works on Mathematics
13. Works on Natural History
14. Popular Narratives

## II TEI MODEL

### 2.1 Rationale for using TEI

We have chosen to adopt the XML standards of the Text Encoding Initiative (TEI) [http7, 2016] as the native data format for *NHSL* primarily for reasons of interoperability and extensibility, as mentioned in our goals above (see Introduction).

First, an XML format is an obvious choice, since XML is used widely and allows for database-like operations on plain-text records in a more-or-less human-readable form. Storing our records natively in XML lets us view, edit, and exchange them with other projects without needing proprietary software. The non-proprietary format also helps ensure the longevity of our data [Heal 2012, 72].

Second, using TEI-XML for *NHSL* engages us in the very large digital humanities community that has already adopted TEI, allowing us to exchange not only data with other projects, but also tools and experiences.

Third, TEI is a format ideally suited to representing information derived from texts. *NHSL* records are "born digital" rather than being a transcription of a printed document; nevertheless, they rely heavily on textual sources, such as manuscript catalogues. Although *NHSL* needs to serve up data to libraries, the XML formats typically used by libraries, such as MODS/MADS and EAD, do not allow us to define the source of specific points of information with the granularity that TEI does. Identifying the sources of the information used to create each record is essential for making *NHSL* useful to the wider scholarly community, not just libraries. Moreover, we plan to write stylesheets that will convert into MODS/MADS or EAD format the essential cataloguing data from our TEI-XML records.

Finally, Syriaca.org is using TEI as the native data format for all of its other projects to date, including, most significantly, the *Digital Catalogue of Syriac Manuscripts in the British Library* [Michelson, forthcoming], which will be a major source for *NHSL*. TEI is particularly well suited to manuscript description [see http8, 2016]. Using TEI across the board makes internal data exchange easier and keeps down costs.

The one major disadvantage of using TEI is that we have not yet discovered other projects that use TEI to describe works in detail, apart from their manifestations in particular manuscripts or editions. Some similar projects are underway using RDF (such as the Perseus Catalog [http9, 2016] which uses the LAWD ontology [http10, 2016), and one goal of our project will be to use TEI in a flexible way that will allow our data to be losslessly serialized into RDF or migrated to that format at a future stage should a standard emerge. In the interim, we are continuing to develop our own strategies reliant on TEI and the depth of markup and vocabulary it provides.

## 2.2 One TEI document per work entity
Each work entity record will consist of a single TEI-XML document containing that work's characteristics and relationships to other entities. This allows for easy editing, convenient and granular git repository management among several team members, and a simple server setup based on the one-to-one correspondence between URIs and XML files. Since we do not want to maintain both sides of work-to-work relationships, these relationships will be stored in only one record for each relationship pair. Each type of relationship will be judged more essential for defining one work entity in the pair than the other, and the relationship will be stored in the record for that work. Thus, for example, works that are derived from other works or from parts of other works will be the ones whose records contain the relationships to such "parent" entities.

## 2.3 TEI document template for works
Each work record will consist of a <bibl> element inside the /TEI/text/body section of the document. The schema for the document is a lightly customized version of the TEI schema, at [http11, 2016]. The /TEI/teiHeader section contains publication information (including editorial responsibility), brief documentation, and a changelog.

The following piece of TEI-XML code shows the author, title, and textLang elements for a sample work with the hypothetical URI http://syriaca.org/work/000.

```
<bibl xml:id="work-000">
    <author ref="http://syriaca.org/person/650" source="#bib000-1 #bib000-3">
        <forename sort="1">Narsai</forename>
    </author>
    <title xml:id="name000-1" xml:lang="syr" source="#bib000-1" syriaca-tags="#syriaca-
        headword">ܣܘܓܝܬܐ ܕܥܠ ܡܠܐܟܐ ܘܡܪܝܡ</title>
    <title xml:id="name000-2" xml:lang="en" source="#bib000-1" syriaca-tags="#syriaca-headword
        #syriaca-anglicized">Sogitha on the Angel & Mary</title>
    <title xml:id="name000-3" xml:lang="syr" source="#bib000-5">ܥܠ ܡܠܐܟܐ ܘܡܪܝܡ ܣܘܓܝܬܐ</title>
    <title xml:id="name000-4" xml:lang="de" source="#bib000-6">Der Engel (Gabriel) und
        Maria</title>
    <title xml:id="name000-5" xml:lang="la" source="#bib000-10">Hymnus de Angelo et
        Maria</title>
    <title xml:id="name000-6" xml:lang="en" source="#bib000-12-1">Canticle on the Annunciation
        of the Virgin</title>
    <title xml:id="name000-7" xml:lang="syr" source="#bib000-11 #bib000-12-1">ܡܐܡܪܐ ܕܥܠ
        ܣܘܓܝܬܐ</title>
    <title xml:id="name000-8" xml:lang="syr-Syrn" source="#bib000-16"> ܣܘܓܝܬܐ ܕܥܠ ܡܠܐܟܐ ܘܡܪܝܡ
        ܘܐܝܙܓܕܐ</title>
    <title xml:id="name000-9" xml:lang="en" source="#bib000-13 #bib000-14 #bib000-12-2
        #bib000-12-3"><foreign xml:lang="syr">ܣܘܓܝܬܐ</foreign> on the Angel and Mary</title>
    <title xml:id="name000-10" xml:lang="en" source="#bib000-15">A Soghitha on the Angel and
        Mary</title>
    <textLang mainLang="syr" source="#bib000-1">Syriac</textLang>
    ...
</bibl>
```

Notes:
- Elements that refer to entities that are not works, publications, or manuscripts contain the entity URI in the ref attribute.
- The source attribute points to one or more <bibl> elements that document the publication or manuscript that provides the basis for an assertion. Since for Syriac works authorship is often uncertain and disputed and titles are unstable, we provide sources even for this type of information that would be unsourced in library catalogs, for instance.[2]
- The record contains multiple titles in multiple languages, each marked with an xml:lang attribute [see http13, 2016].[3] While we do not designate any titles as authoritative, we do label up to one standardized title per language with the attribute syriaca-tags="#syriaca-headword" [see http14, 2016]. These headwords are visualized as record titles on Syriaca.org and may be used by library catalogs to construct uniform titles. However, the headwords are not unique identifiers; only the URI should be considered a unique identifier.

We use <note> elements with several different type attributes to provide additional identifying information about the work.

---

[2] The source attribute is not yet allowed on author, title, and a number of other elements. However, it will be made a global attribute in the next TEI release [http12, 2016; Cayless, 2016].

[3] Syriaca.org uses the ISO 639 Syriac macrolanguage code "syr" for all Syriac, despite the fact that there is a separate and, so far, unrelated ISO 639 code "syc" for "Classical Syriac." This is to better accommodate current practices by major sites such as WorldCat and to avoid making potentially artificial distinctions between "classical" and other varieties of Syriac. We have petitioned the ISO administrator to include "syc" under the "syr" macrolanguage.

- <note type="abstract"> provides a short prose description of the work.
- <note type="prologue">, <note type="incipit">, and <note type="explicit"> give excerpts from the work to help recognize it in manuscripts and disambiguate it from other works with the same title. Notes indicate the resource that provided the excerpt using the source attribute, and excerpts are embedded in <quote> tags.
- <note type="disambiguation"> contains any additional discussion (when needed) about how to distinguish the work from other works with which it may be confused.

If the text of such notes is available in more than one language, each language version is contained in a <seg> element marked with an xml:lang attribute.

The TEI <idno> element is where we declare the Syriaca.org URI of the work entity being described and match it to any identifiers in other systems. In the following example, the external identifiers are reference numbers from two earlier reference works, *Bibliotheca Hagiographica Syriaca* (*BHS*) and *Bibliotheca Hagiographica Orientalis* (*BHO*):

```
<idno type="URI">http://syriaca.org/work/270</idno>
<idno type="BHS">49</idno>
<idno type="BHO">772</idno>
```

Inside the <bibl> node describing the work entity, we enclose manuscripts and publications in additional <bibl> tags with type attributes specifying a prefixed RDF class to which they belong (either lawd:WrittenWork [http15, 2016] or one of its subclasses). Since the complete bibliographic record is stored at the URI designated in ptr/@target, there is no need to give full bibliographical details here — only the <citedRange> and some basic information to enable the work entity TEI record to be human readable and be sufficient without recourse to the complete bibliographic record. We are investigating how best to link to these <bibl> elements within the TEI document in a way that references both the URI of the bibliographic record and the specific cited range to which we are referring. Options we are considering include using data pointers to the @xml:id of the <bibl> element (for example, "#bib270-4") [http16, 2016], using the Canonical Text Service [http17, 2016], or minting more granular URIs for each section of the publication.

The following is an example of a reference to a publication (using a hypothetical URI):

```
<bibl type="lawd:Edition" xml:id="bib270-4">
    <author>
        <forename>P.</forename>
        <surname>Bedjan</surname>
    </author>
    <title level="m" xml:lang="la">Acta Martyrum et Sanctorum</title>
    <ptr target="http://syriaca.org/bibl/10001"/>
    <citedRange unit="volume" from="2" to="2">2</citedRange>
    <citedRange unit="pp" from="260" to="275">260-275</citedRange>
</bibl>
```

We mark up combined edition-translations (which are quite common) using multiple <bibl> nodes, one for each language, each with its own @xml:id. In such cases, each <bibl> includes a reference to the specific pages or columns for that language, as well as a <textLang> element. This allows us to separate the work-edition relationship from the work-translation relationship.

The following is a reference to a manuscript (again with a hypothetical URI):

```xml
<bibl type="lawd:WrittenWork" xml:id="bib270-6">
    <msIdentifier>
        <country>Germany</country>
        <settlement>Berlin</settlement>
        <collection xml:lang="de">Königliche Bibliothek</collection>
        <idno type="URI">http://syriaca.org/manuscript/20001</idno>
        <altIdentifier>
            <idno type="KB-Shelfmark">or. oct. 1257</idno>
        </altIdentifier>
    </msIdentifier>
    <biblScope>
        <locus from="1" to="23">1-23</locus>
        <idno type="URI">http://syriaca.org/manuscript/20001#a1</idno>
    </biblScope>
</bibl>
```

Note that the URI contained in the biblScope/idno refers to the specific section of the manuscript record that describes the locus where this work appears (1-23).

Finally, we use a <listRelation> block to specify the connections between the work and all other entities. As recommended by [Jordanous et al, 2012, §2.2.1] of the Sharing Ancient Wisdoms (SAWS) project [http18, 2016], we are modeling RDF triples in the <relation> element, with @active as the Subject, @ref as the Predicate, and @passive as the Object. For example,

```xml
<listRelation>
    <relation type="editions" active="#bib270-4 #bib270-5" ref="lawd:embodies"
        passive="http://syriaca.org/work/270" source="#bib270-1"/>
    <relation type="ancientVersion" active="#bib270-15" ref="lawd:embodies"
        passive="http://syriaca.org/work/270" source="#bib270-1"/>
    <relation type="mss" active="http://syriaca.org/manuscript/20001#a1
        http://syriaca.org/manuscript/20002#b1" ref="lawd:embodies"
        passive="http://syriaca.org/work/270" source="#bib270-1"/>
    <relation type="mssWitnesses" active="#bib270-4" ref="dct:source"
        passive="http://syriaca.org/manuscript/20001#a1"/>
    <relation type="modernTranslation" active="#bib270-16" ref="dct:source"
        passive="http://syriaca.org/work/270" source="#bib270-1"/>
    <relation ref="syriaca:commemorates" active="http://syriaca.org/work/270"
        passive="http://syriaca.org/person/1922 http://syriaca.org/person/1544
        http://syriaca.org/person/1383" source="#bib270-1"/>
</listRelation>
```

## III PROJECT STATUS AND NEXT STEPS

As of March 2016, the editors are continuing to develop the data model for *NHSL* and are interested in feedback regarding this model. One of the remaining steps is developing a detailed TEI to RDF crosswalk for Syriac works so that we can represent these records as linked open data. We believe that serializing these documents as RDF will allow us to better aggregate data and improve semantic awareness. For this reason, we are exploring several vocabularies, including Linking Ancient World Data (LAWD) [http10, 2016], which is designed specifically for enriching existing vocabulary schemes to suit the needs of projects relating to the ancient world.

The next step will be extracting data from a variety of resources, primarily manuscript catalogues. Starting from manuscript catalogues maintains the connection between work entities and manuscript representations and allows for a fresh, ground-up approach that brings to light even works that have otherwise been neglected. Syriaca.org is already encoding in TEI one of the most important printed manuscript catalogues (over 1,000 manuscripts) [Wright, 1870-1872; Michelson, forthcoming] and has acquired data from partners for over 5,000 additional manuscripts. These and other sources will provide the core data for producing records for Syriac literary works.

Next, we will need to disambiguate the Syriac works generated from this core data. Given the instability of both titles and author attributions for Syriac works, record linkage will require heavy editorial involvement and non-traditional approaches. Loading the records into an XML database and querying with xQuery or natural language processing tools may help the editors discover records to be matched. These disambiguated records will be the basis for canonical URI's of Syriac works.

The final step will be presenting the records in a searchable, human-readable form. Syriaca.org has been using eXistdb with HTML front-ends to serve up other XML datasets and already has a prototype visualization for Syriac work records. Development is also underway on an XForms-based editorial module that will accept user submissions with editorial review.

## Conclusion

*The New Handbook of Syriac Literature* will be an open-access, scholarly resource that establishes canonical URIs for Syriac works and serves as a guide to the embodiments of those works in manuscripts and publications. Work records will be encoded in TEI-XML, and relationships to other entities will use RDF properties so that they may be easily serialized into RDF formats. *NHSL* will integrate closely with other Syriaca.org resources and will use linked open data methods to connect with other projects and resources.

We welcome comments and collaboration not only from scholars within Syriac studies but also from the emerging community of digital scholars engaged in the study of corpora in ancient languages.